\documentclass[reprint, aip,  letterpaper, superscriptaddress]{revtex4-2}

\usepackage[T1]{fontenc} 

\usepackage{amsmath,color}
\usepackage{amssymb}
\usepackage{amsfonts}
\usepackage{amsthm}
\usepackage{mathtools}

\usepackage[title]{appendix}

\usepackage{algorithm}
\usepackage{algpseudocode}
\usepackage{dsfont}

\usepackage{bbold}
\usepackage{chemformula}
\usepackage{siunitx}
\usepackage{braket}

\usepackage{accents}
\newcommand{\dbtilde}[1]{\accentset{\approx}{#1}}

\newcommand{\vE}{\mathbf{E}}

\newcommand{\ve}{\mathbf{e}}

\newcommand{\vmu}{\boldsymbol{\mu}}

\newcommand{\vxi}{\boldsymbol{\xi}}

\newcommand{\hH}{\hat{H}}

\newcommand{\hV}{\hat{V}}

\newcommand{\avg}[1]{\left\langle #1\right\rangle}
\newcommand{\red}[1]{{\color{black} #1}}

\begin{document}

	\title{Polariton relaxation under vibrational strong coupling: Comparing cavity molecular dynamics simulations against Fermi's golden rule rate}
	
	\author{Tao E. Li}%
	\email{taoli@sas.upenn.edu}
	\affiliation{Department of Chemistry, University of Pennsylvania, Philadelphia, Pennsylvania 19104, USA}

	\author{Abraham Nitzan} 
	\email{anitzan@sas.upenn.edu}
	\affiliation{Department of Chemistry, University of Pennsylvania, Philadelphia, Pennsylvania 19104, USA}
	\affiliation{School of Chemistry, Tel Aviv University, Tel Aviv 69978, Israel}

	\author{Joseph E. Subotnik}
	\email{subotnik@sas.upenn.edu}
	\affiliation{Department of Chemistry, University of Pennsylvania, Philadelphia, Pennsylvania 19104, USA}
	
	\begin{abstract}
	Under vibrational strong coupling (VSC), the formation of molecular  polaritons may significantly modify the photo-induced or thermal properties of molecules. In an effort to understand these intriguing modifications, both experimental and theoretical studies have focused on the ultrafast dynamics of vibrational polaritons. Here, following our recent work [J. Chem. Phys., \textbf{154}, 094124, (2021)], we systematically study the mechanism of  polariton relaxation for liquid \ch{CO2} under a weak external pumping.  Classical cavity molecular dynamics (CavMD) simulations confirm that  polariton relaxation results from the combined effects of (i) cavity loss through the photonic component and (ii) dephasing of the bright-mode component to vibrational dark modes as mediated by intermolecular interactions. The latter polaritonic dephasing rate is proportional to the product of the weight of the bright mode in the polariton wave function and the spectral overlap between the polariton and dark modes. Both these factors are sensitive to  parameters such as the Rabi splitting and cavity mode detuning. Compared to a Fermi's golden rule calculation based on a tight-binding harmonic model, CavMD yields a similar parameter dependence for the upper polariton relaxation lifetime but  sometimes a modest disagreement  for the lower polariton. We suggest that this  disagreement results from polariton-enhanced molecular nonlinear absorption due to molecular anharmonicity, which is not included in our analytical model. We also summarize recent progress on probing nonreactive VSC dynamics with CavMD.
	\end{abstract}

	\maketitle
	
	\section{Introduction}
	
	Strong light-matter interactions provide a novel approach towards modifying molecular properties by forming hybrid light-matter states --- known as  polaritons \cite{Flick2017,Ribeiro2018,Flick2018,Herrera2019,Hirai2020review,Climent2021,Xiang2021JCP,Garcia-Vidal2021}. In the past decade, vibrational strong coupling (VSC) was experimentally observed when a large number of molecules in the condensed phase were placed in a Fabry--P\'erot microcavity and a vibrational mode of molecules formed strong coupling with a cavity mode \cite{Shalabney2015,Long2015,Simpkins2015,George2015}. Under VSC, one of the most intriguing experimental observations is the possibility of modifying thermally activated ground-state chemical reactions  through such a Fabry--P\'erot microcavity and without a direct external pumping \cite{Thomas2016}.
	Inspired by this seminal observation \cite{Thomas2016} from the Ebbesen group, intensive experimental and theoretical efforts have now been made to understand VSC-related chemistry. From an experimental point of view, there has been an extensive exploration of VSC catalytic effects in different types of chemical reactions \cite{Vergauwe2019,Thomas2019,Thomas2019_science,Lather2019,Pang2020,Sau2021,Lather2021} and the use of VSC to alter other molecular properties in the absence of an external laser pumping \cite{Hirai2020Crys} has also been investigated; at the same time, spectroscopists have also focused on the ultrafast dynamics of vibrational polaritons by pump-probe \cite{Dunkelberger2016,Dunkelberger2018,Dunkelberger2019} and two-dimensional infrared (IR) \cite{Xiang2018,Xiang2019Nonlinear,Xiang2019State,Xiang2020Science,Grafton2021,Xiang2021} spectroscopies. From a theoretical point of view, while the detailed mechanism of "VSC catalysis" is still not well understood \cite{Galego2019,Campos-Gonzalez-Angulo2019,Li2020Origin,Campos-Gonzalez-Angulo2020,LiHuo2020,Climent2020,Sidler2021,DuNoneq2021,Schafer2021}, the current models of  ultrafast polariton dynamics appear to be largely consistent with observations \cite{Pino2015,Saurabh2016,F.Ribeiro2018,Du2018,Zhang2019,Ribeiro2020,Li2020Nonlinear}.
	
	As far as the ultrafast dynamics of vibrational polaritons are concerned, recent experiments \cite{Dunkelberger2016,Dunkelberger2018,Xiang2018,Xiang2019Nonlinear,Xiang2019State,Xiang2020Science,Grafton2021}  have emphasized the complicated interaction between polaritons  and vibrational dark modes after the polariton pumping with an external laser field; see Ref. \cite{Xiang2021JCP} for a recent review. Two of the most  nontrivial observations include (i) polariton mediated intermolecular vibrational energy transfer between two molecular species after the upper polariton (UP) pumping \cite{Xiang2020Science} and (ii) polariton enhanced molecular nonlinear absorption after the lower polariton (LP) pumping \cite{Xiang2019State,Xiang2021JCP}. While there  have been a few theory papers  on polariton relaxation based on analytic models \cite{Pino2015,Du2018,Saez-Blazquez2018,Sommer2021,Engelhardt2021} and realistic modeling \cite{Luk2017,Groenhof2019,Tichauer2021,Triana2020}, a systematic  study of  vibrational polariton relaxation when realistic molecules are considered is still unavailable, which will be the focus of this paper.

	Our approach for studying polariton relaxation is classical cavity molecular dynamics (CavMD) simulations \cite{Li2020Water,Li2020Nonlinear,Li2021Collective,Li2021Solute}, in which both the cavity modes and molecular vibrations are treated classically and move along an electronic ground state surface. Compared with standard theories for strong light-matter interactions, CavMD  has two advantages:  the ability to (i) describe realistic molecules and (ii)  treat a very large number of molecules coupled to the cavity with an affordable computational cost. These advantages allow CavMD to identify nontrivial VSC effects beyond, e.g., the Tavis--Cummings Hamiltonian \cite{Tavis1968,Tavis1969} or the coupled oscillator model \cite{Rudin1999,Junginger2020}. The latter is  necessary because, to date, most VSC experiments were performed with Fabry--P\'erot cavities, in which the cavity volume is $\sim \lambda^3$ (where $\lambda$ denotes the wavelength of the cavity mode) and the corresponding molecular number is very large (say, $N \sim 10^{10}$).
	
	\begin{figure*}
		\centering
		\includegraphics[width=0.8\linewidth]{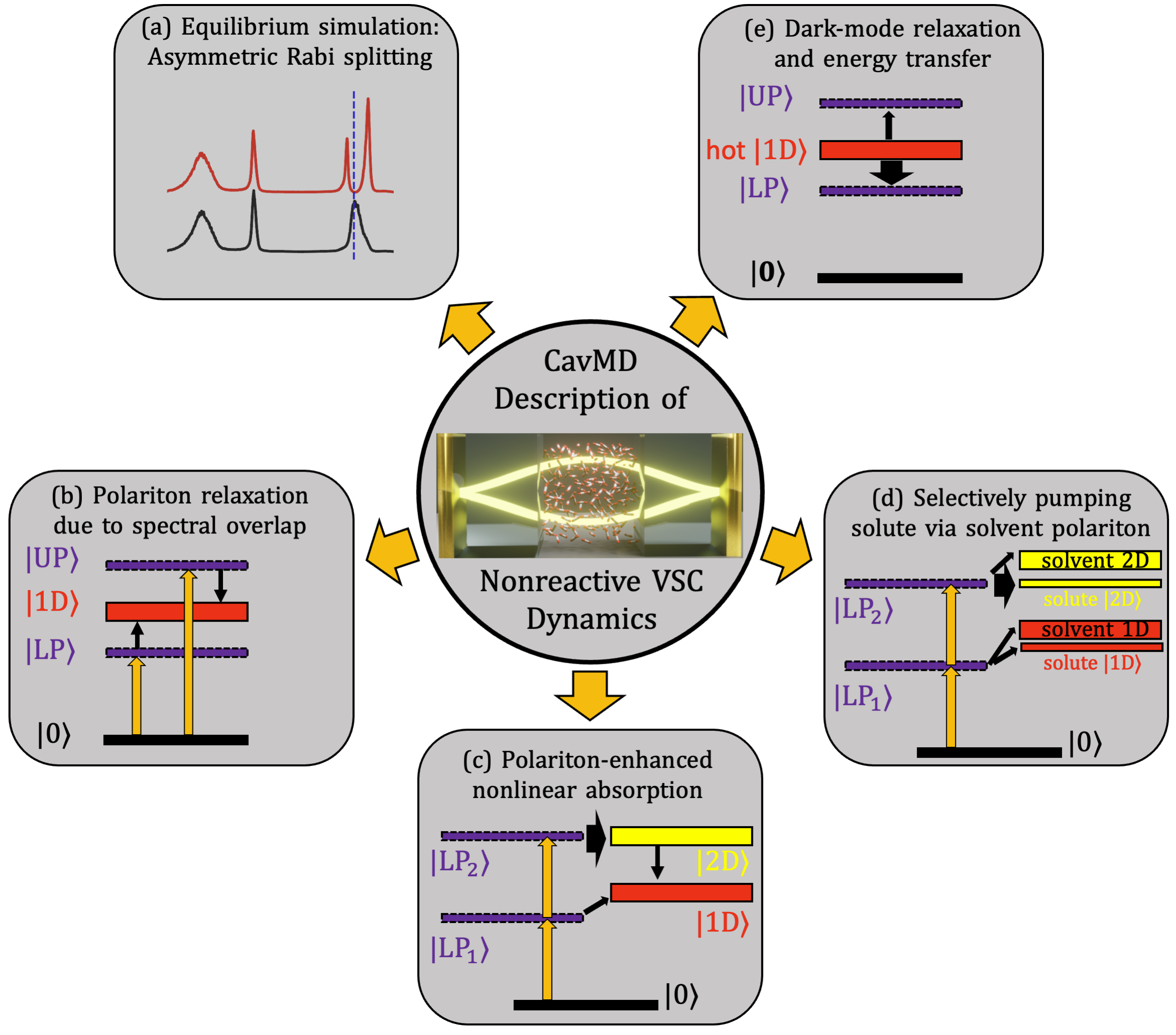}
		\caption{Illustration of  nonreactive VSC dynamics probed by CavMD simulations. (a) Asymmetric Rabi splitting under thermal equilibrium \cite{Li2020Water}. (b) Polariton relaxation to dark modes (this work). (c) Polariton-enhanced molecular nonlinear absorption \cite{Li2020Nonlinear}. (d) Selectively exciting solute molecules via solvent polariton pumping \cite{Li2021Solute}. (e) Dark-mode relaxation and energy transfer dynamics \cite{Li2021Collective}.  The cartoon in the middle circle demonstrates the simulation setup of CavMD, which is adapted from Ref. \cite{Li2020Nonlinear}.}
		\label{fig:CavMD_summary}
	\end{figure*} 
	
	We note that recent considerations of the role of VSC played by symmetry and selection rules can in principle be studied by CavMD. However, we do not address these issues here. We also emphasize that CavMD is a classical approximation, so this approach may fail if any quantum effect (either from the nuclei or cavity photons) dominates the VSC effect (which is  largely unexplored). In order to understand the applicability of CavMD,  it is necessary to benchmark the performance of CavMD under various situations against experiments or analytic models; see the next paragraph and also  Fig. \ref{fig:CavMD_summary} for a summary. Following the method in Ref. \cite{Li2020Nonlinear}, here we will focus on using CavMD to describe polariton relaxation  when the polariton is weakly pumped by an external laser field. In the weak pumping limit (which is also the condition for most ultrafast experiments), we will compare the CavMD results with a quantum model where the polariton relaxation lifetime can be calculated by a Fermi's golden rule (FGR) rate. Together with an extensive  parameter dependence study of polariton relaxation, this comparison will not only serve as an important performance benchmark of CavMD, but also facilitate a mechanistic understanding of vibrational polariton relaxation.

	Before focusing on the details of vibrational polariton relaxation, as shown in Fig. \ref{fig:CavMD_summary}, let us summarize some of the most important observations by CavMD from a series of recent publications (including this work) as well as the connection to different experiments and theories.
	\begin{itemize}
		\item[(a)] Asymmetric Rabi splitting from equilibrium CavMD simulations \cite{Li2020Water}. 
		We have included the self-dipole term in CavMD, \footnote{Note that there is a debate on the proper gauge-invariant quantum-electrodynamical Hamiltonian for strong coupling \cite{DiStefano2019,Schafer2020,Taylor2020,Andrews2018,Feist2020}}\textsuperscript{,} \cite{DiStefano2019,Schafer2020,Taylor2020,Andrews2018,Feist2020} a direct consequence of which is a renormalized vibrational frequency inside the cavity \cite{Cortese2021} and the asymmetry of the Rabi splitting even when the cavity mode is at resonance with the bare vibrational mode outside the cavity. For the linear IR spectrum of liquid water under VSC, CavMD not only yields similar results as the coupled oscillator model, but also shows that the polaritonic lineshape is much smaller than the inhomogeneous broadening of the water \ch{O-H} peak. This result --- which cannot be obtained from the coupled oscillator model ---  agrees with an early theoretical prediction \cite{Houdre1996} and experimental observations for liquid water VSC \cite{Vergauwe2019}. This simulation also suggests that individual molecular properties are not meaningfully modified under VSC when thermal equilibrium is considered, suggesting  a nonequilibrium or quantum origin of "VSC catalysis".
		
		\item[(b)] Mechanism of vibrational polariton relaxation (this work). Here, we will demonstrate that, for the UP,  CavMD predicts a similar parameter dependence  for the UP lifetime as does a FGR  rate calculation based on a tight-binding harmonic model. We also highlight the significance of the  dephasing process from the polariton to the dark modes due to the spectral overlap between the polariton and dark modes; this overlap can be controlled by the Rabi splitting and cavity mode detuning. 
		The results here validates the ability	of the CavMD simulation approach as far as the capacity to match more conventional quantum optics approaches in polariton physics.
		
		\item[(c)] Polariton enhanced molecular nonlinear absorption \cite{Li2020Nonlinear}. When twice the LP roughly matches the vibrational $0\rightarrow2$ transition (due to molecular anharmonicity), CavMD shows that strongly exciting the LP can directly transfer energy to the second  ($\ket{2\text{D}}$) or higher excited states of vibrational dark modes; at later times, this excess energy gradually transfers to the first excited state of vibrational dark modes ($\ket{1\text{D}}$). The delayed excitation of $\ket{1\text{D}}$ states after the LP pumping has been experimentally reported previously \cite{Xiang2019State} and a direct observation of the initial pumping of  $\ket{2\text{D}}$ states has also been shown recently \cite{Xiang2021JCP} . These experimental observations, together with a quantum model study \cite{Ribeiro2020}, appear to validate our simulation results.
		
		\item[(d)] Selectively exciting solute molecules via solvent polariton pumping \cite{Li2021Solute}. Given the validity of the polariton enhanced molecular nonlinear absorption mechanism, CavMD shows that, if the LP of the solvent molecules can support the nonlinear transition of the solute (not the solvent) molecules only (i.e., twice the solvent LP roughly matches the solute $0\rightarrow2 $ vibrational transition), pumping the solvent LP may highly excite the solute molecules, leaving the solvent molecules barely excited. This finding deserves experimental exploration in the near future.
		
		\item[(e)] Nonequilibrium dark-mode dynamics \cite{Li2021Collective}. For a small fraction of hot \ch{CO2} molecules immersed inside a \ch{CO2} thermal bath, in the absence of external pumping, the LP can be transiently excited even when the molecular system size is large. The transiently excited LP can accelerate vibrational relaxation of the hot molecules and promote the energy transfer to the thermal molecules. For small-volume cavities ($N < 10^4$), the VSC effect on the relaxation rate per molecule resonantly depends on the cavity mode frequency; for cavities with larger volumes, the VSC effect per molecule vanishes, despite the fact that the transiently excited LP prefers to transfer energy to the molecules at the tail of the energy distribution (which may connect to a VSC  catalytic effect). The fact that all VSC effects on dark mode relaxation become vanishingly small (per molecule) for a large cavity volume is consistent with a recent experiment with Fabry--P\'erot cavities.\cite{Grafton2021}
	\end{itemize}
	
	We now turn to (b) above and in the rest of this paper focus on investigating the mechanism of vibrational polariton relaxation.
	This paper is organized as follows. In Sec. \ref{sec:methods}, we introduce  the FGR calculation and CavMD simulations that we will employ in order to describe the polariton relaxation lifetime. In Sec. \ref{sec:results} we present results. We conclude in Sec. \ref{sec:conclusion}.
	
	\section{Methods}\label{sec:methods}
	
	\subsection{Model system and the FGR rate}
	
	The standard theory for describing strong light-matter interactions in the collective regime is the Tavis--Cummings (TC) Hamiltonian \cite{Tavis1968,Tavis1969}. In this model, a single cavity mode with frequency $\omega_c$  is coupled to $N$ identical two-level systems with transition frequency $\omega_0$:
	\begin{equation}\label{eq:H_TC}
		\hH = \hbar\omega_c \hat{a}^{\dagger}\hat{a} 
		+ \hbar\omega_0\sum_{n=1}^{N}\hat{\sigma}_{+}^{(n)}
		\hat{\sigma}_{-}^{(n)}
		+ \hbar g_0 \sum_{n=1}^{N}\left(
		\hat{a}^{\dagger}\hat{\sigma}_{-}^{(n)} + \hat{a}\hat{\sigma}_{+}^{(n)}
		\right) .
	\end{equation}
	Here, $\hat{a}^{\dagger}$ and $\hat{a}$ denote the creation and annihilation operators for the cavity photon; $\hat{\sigma}_{+}^{(n)} \equiv \ket{ne}\bra{ng}$ and $\hat{\sigma}_{-}^{(n)} \equiv \ket{ng}\bra{ne}$, where $\ket{ng}$ and $\ket{ne}$ denote the ground and excited state for the $n$-th two-level system; $g_0$ denotes the coupling constant between the cavity photon and each two-level system. In the light-matter coupling term (the last term above), both the long-wave approximation and the rotating-wave approximation have been taken.
	
	For VSC, since high-frequency molecular vibrations can  be approximated as harmonic oscillators at and below room temperature, we  replace the two-level systems in Eq. \eqref{eq:H_TC} by quantum harmonic oscillators (with the creation and annihilation operators denoted by $\hat{b}_{n}^{\dagger}$ and $\hat{b}_{n}$) and obtain the following Hamiltonian:
	\begin{equation}\label{eq:H_TC_harm}
		\hH= \hbar\omega_c \hat{a}^{\dagger}\hat{a} 
		+ \hbar\omega_0\sum_{n=1}^{N}\hat{b}_{n}^{\dagger} \hat{b}_{n}
		+ \hbar g_0 \sum_{n=1}^{N}\left(
		\hat{a}^{\dagger}\hat{b}_{n} + \hat{a}\hat{b}_{n}^{\dagger}
		\right) .
	\end{equation}
    This model Hamiltonian serves as the starting point of our derivation. 
    
    In the light-matter coupling term of Eq. \eqref{eq:H_TC_harm}, since the cavity photon interacts  with a symmetric combination of the molecular operators only, one can introduce the bright-mode creation and annihilation operators as follows:
	\begin{equation}\label{eq:B}
	    \begin{aligned}
		\hat{B} &= \frac{1}{\sqrt{N}} \sum_{n=1}^{N} \hat{b}_{n} , \\
		\hat{B}^{\dagger} & = \frac{1}{\sqrt{N}} \sum_{n=1}^{N} \hat{b}_{n}^{\dagger} .
		\end{aligned}
	\end{equation}
	All the other $N-1$  linear combinations of the molecular operators are decoupled from the cavity mode and are called the dark modes:
	\begin{equation}\label{eq:Dmu}
	\begin{aligned}
		\hat{D}_{\mu} &= \frac{1}{\sqrt{N}} \sum_{n=1}^{N} e^{i2\pi n \mu / N} \hat{b}_{n} , \\
		\hat{D}_{\mu}^{\dagger} &= \frac{1}{\sqrt{N}} \sum_{n=1}^{N} e^{i2\pi n \mu / N} \hat{b}_{n}^{\dagger} .
	\end{aligned}
	\end{equation}
	Here, $\mu = 1, 2, \cdots, N-1$ indexes the dark modes.
	
	With the separation of bright and dark modes, the Hamiltonian in Eq. \eqref{eq:H_TC_harm} becomes
	\begin{equation}\label{eq:H_TC_separated}
		\hH = \hbar\omega_c \hat{a}^{\dagger}\hat{a} 
		+ \hbar\omega_0\hat{B}^{\dagger}\hat{B}
		+ \frac{1}{2}\hbar\Omega_N \left(\hat{a}^{\dagger}\hat{B} + \hat{a}\hat{B}^{\dagger}\right) + \hH_{\text{D}} ,
	\end{equation}
	where $\Omega_N \equiv 2\sqrt{N}g_0$ and $\hH_{\text{D}}$ is the Hamiltonian for the dark modes:
	$\hH_{\text{D}} = \hbar\omega_0\sum_{\mu=1}^{N-1}\hat{D}_{\mu}^{\dagger} \hat{D}_{\mu}$.
	Eq. \eqref{eq:H_TC_separated} can be further diagonalized, which leads to the following Hamiltonian:
	\begin{equation}\label{eq:H_TC_eigen}
		\hH = \hbar\omega_{+} \hat{P}_{+}^{\dagger}\hat{P}_{+} 
		+ \hbar\omega_{-} \hat{P}_{-}^{\dagger}\hat{P}_{-} 
		+ \hH_{\text{D}} .
	\end{equation}
	Here, the frequencies of the eignestates are
	\begin{equation}\label{eq:LP_UP_freq}
		\omega_{\pm} = \frac{1}{2}\left[\omega_0 + \omega_c \pm \sqrt{\Omega_N^2 + (\omega_0- \omega_c)^2}\right] .
	\end{equation}
	When the cavity photon and the bright mode are at resonance ($\omega_c = \omega_0$), the frequency difference between the two eigenstates is $\omega_{+} - \omega_{-} = \Omega_N$. Once $\Omega_N$ exceeds the linewidth of the bright mode (which can be measured from the linear spectrum of molecules outside the cavity) and the loss rate of the cavity (due to the imperfections in the cavity mirrors), a peak splitting can be measured from the linear spectrum of the coupled cavity-molecular system. The peak splitting, which is called the collective Rabi splitting, is equal to $\Omega_N = 2\sqrt{N} g_0$.
	
	In Eq. \eqref{eq:H_TC_eigen}, the creation and annihilation operators for the eigenstates ($\hat{P}_{\pm}^{\dagger}$ and $\hat{P}_{\pm}$) are defined as linear combinations of the cavity-photon and bright-mode operators:
	\begin{equation}\label{eq:Ppm}
	\begin{aligned}
		\hat{P}_{\pm} &= X_{\pm}^{(\text{B})} \hat{B}  + X_{\pm}^{(\text{c})} \hat{a},  \\
		\hat{P}_{\pm}^{\dagger} &= X_{\pm}^{(\text{B})} \hat{B}^{\dagger}   + X_{\pm}^{(\text{c})} \hat{a}^{\dagger},
	\end{aligned}
	\end{equation}
	where the coefficients are $X_{+}^{(\text{B})} = -X_{-}^{(\text{c})}  = -\sin(\theta)$, $X_{-}^{(\text{B})} = X_{+}^{(\text{c})}  = \cos(\theta)$, and the mixing angle $\theta$ is $\theta = \frac{1}{2} \tan^{-1}\left(\frac{\Omega_N}{\omega_c - \omega_0}\right)$.
	When a Rabi splitting is observed, the two eigenstates ($-$ and $+$) are called the lower (LP) or upper (UP) polariton, respectively.
	
	\subsubsection{Calculating the polariton lifetime}

	In the absence of intermolecular interactions, as shown in Eq. \eqref{eq:H_TC_eigen}, the polaritons are completely decoupled from the dark modes, so once the LP or UP is excited to the first excited state,  the polaritonic decay rate  ($1/\tau_{\pm}$) should be simply  a linear combination of the relaxation rates of the photonic and the molecular parts \cite{Deng2010} \footnote{Note that this expression is valid only when Rabi splitting is larger than the linewidth of either the photonic or molecular peak}:
	\begin{equation}\label{eq:polariton_rate_simple}
		\frac{1}{\tau_{\pm}} =    |X_{\pm}^{(\text{c})}|^2 k_{c} + |X_{\pm}^{(\text{B})}|^2 k_{\text{B}},
	\end{equation}
	where $k_{c}$ denotes the cavity loss rate, and $k_{\text{B}}$ denotes the relaxation rate of the bright state from the first excited to the ground state outside a cavity, which could be approximated as the linewidth of the vibrational peak from molecular linear IR spectroscopy outside the cavity  (if the molecular system has no inhomogeneous broadening). Note that the bright mode relaxation rate $k_\text{B}$ reflects both the radiative and non-radiative relaxation pathways, and the radiative relaxation contribution can be superradiant (meaning a $N$ dependence). For molecular vibrations, since the radiative relaxation rates (with a typical lifetime of seconds) are usually much slower than the non-radiative rates (with a typical lifetime from ns to ps), we will disregard the radiative relaxation contribution of $k_\text{B}$. \footnote{When considering the radiative relaxation of $k_{\text{B}}$, we also disregard the possibility that collective radiative emission can be faster than single molecule emission and also the Purcell effect.}
	
	The rate equation in Eq. \eqref{eq:polariton_rate_simple} has been widely known by the polariton community.
	For realistic molecular systems in the condensed phase, however, we need to consider another decay pathway of the polaritons. Since molecules do interact with each other, polaritons (which are partially composed the molecular degrees of freedom) can interact with the dark modes through intermolecular interactions. Naturally, the interaction between polaritons and dark modes may introduce an additional dephasing pathway for the decay of polaritons.
	
	In order to account for this additional dephasing process, on top of the standard polaritonic Hamiltonian in Eq. \eqref{eq:H_TC_eigen}, we add a simple tight-binding intermolecular interaction between neighboring molecules:
	\begin{equation}\label{eq:V_tb}
		\hV = \sum_{n=1}^{N} \hbar \Delta_n\left[\sum_{M_n=1}^{N_{\text{nn}}}\left(\hat{b}_{n}^{\dagger} \hat{b}_{M_n} 
		+ \hat{b}_{n} \hat{b}_{M_n}^{\dagger}
		\right)\right],
	\end{equation}
	where $M_n$ indices all the possible nearest neighbors (with a total number of $N_{\text{nn}}$) of molecule $n$,  and $\Delta_n$ denotes the coupling strength between the nearest neighboring molecules. For different $n$, $\{\Delta_n\}$ are uncorrelated but should have similar magnitude.  In Eq. \eqref{eq:V_tb}, since the operators of each molecule can be expressed as linear combinations of the bright and dark modes, the interaction between bright (which contributes to the polaritons) and dark modes has been included. In order to better illustrate this point, if we consider only the bright-mode contribution of $\hat{b}_{M_n}$, i.e., $\hat{b}_{M_n} \leftarrow \hat{B} / \sqrt{N} + \cdots$,
	we can approximate Eq. \eqref{eq:V_tb} as
	\begin{equation}\label{eq:V_transform}
		\begin{aligned}
			\hV &= \frac{\hbar N_{\text{nn}}}{\sqrt{N}}\sum_{n=1}^{N} \Delta_n \left(
			\hat{b}_{n}^{\dagger}\hat{B} + \hat{b}_{n} \hat{B}^{\dagger}
			\right) + \cdots,
		\end{aligned}
	\end{equation}
	where $\cdots$ represents the interaction between individual molecules (indexed by $n$) and the dark modes.
	For large $N$, since $\hat{b}_{n}$ ($\hat{b}_{n}^{\dagger}$) is predominately composed of the dark modes, Eq. \eqref{eq:V_transform} can be used to calculate the polaritonic dephasing rate to the dark modes.
	
	We now calculate the dephasing rate from polaritons to vibrational dark modes with a FGR calculation:
	\begin{subequations}
		\begin{eqnarray}
			\gamma = \sum_{f}\frac{2\pi}{\hbar^2} |V_{fi}|^2 \delta(\omega - \omega_f).
		\end{eqnarray}
	\end{subequations}
	Here $i$ and $f$ denotes the initial (polaritonic) and final (dark) states; $\delta(\omega - \omega_f)$ denotes the density of states (DOS) for state $f$;  $V_{fi} = \braket{f|\hat{V}|i}$ denotes the transition matrix element, where we have defined $\hat{V}$ in \eqref{eq:V_transform}.
	By substituting $\ket{i} = \hat{P}_{\pm}^{\dagger}\ket{0}$ (exciting a polariton) and the definition of $\hat{P}_{\pm}^{\dagger}$ in Eq. \eqref{eq:Ppm}, we obtain the transition matrix element as: 
	\begin{equation}
			V_{fi} = \braket{f|\hat{V}|i} \approx \frac{\hbar N_{\text{nn}}\Delta_f}{\sqrt{N}} X^{(\text{B})}_{\pm}.
	\end{equation}
	Hence, the polaritonic dephasing rate becomes
	\begin{equation}\label{eq:gamma_pre}
	\begin{aligned}
	    \gamma_{\pm} &\approx \sum_{f=1}^{N} \frac{2\pi}{\hbar^2} \left( \frac{\hbar N_{\text{nn}}\Delta_f}{\sqrt{N}} X_{\pm}^{(\text{B})} \right)^2  \delta(\omega - \omega_f), \\
	    &=   2\pi \Delta^2 |X_{\pm}^{(\text{B})}|^2 \rho_0(\omega), 
	\end{aligned}
	\end{equation}
 	Here, we have  defined $\Delta^2 \equiv \sum_{f=1}^{N} N_{\text{nn}}^2 \Delta^2_f \delta(\omega-\omega_f)  / \sum_{f=1}^{N}\delta(\omega-\omega_f)$ to denote the average intermolecular coupling strength; $\rho_0(\omega) \equiv \frac{1}{N} \sum_{f=1}^{N}\delta(\omega-\omega_f)$ denotes the DOS for the dark-mode manifold per molecule. Note that here we have assumed that polariton states are orthogonal to the individual molecular basis ($\{\hat{b}_n\}$), which is valid in the limit of large $N$.
	
	In the presence of disorder, the polariton is not an eigenstate of the Hamiltonian and has also a finite experimental linewidth. In our calculation we introduce a
	phenomenological DOS for the polariton to account for this width
	 ($\rho_{\pm}(\omega)$, where $\int_{-\infty}^{\infty} d\omega \rho_{\pm}(\omega) = 1$), so the polaritonic dephasing rate is further smeared out:
	\begin{subequations}\label{eq:gamma_pm}
	\begin{equation}
	    \gamma_{\pm} = 2\pi \Delta^2|X_{\pm}^{(\text{B})}|^2 J_{\pm},
	\end{equation}
	where $J_{\pm}$ denotes the spectral overlap between the polariton and the dark modes:
	\begin{equation}\label{eq:Jpm}
	    J_{\pm} \equiv \int_{0}^{+\infty} d\omega \rho_0(\omega) \rho_{\pm}(\omega).
	\end{equation}
	\end{subequations}
	Eq. \eqref{eq:gamma_pm} shows that the polaritonic dephasing rate is determined by three factors: the magnitude of intermolecular interactions ($\Delta^2$), the molecular weight of polaritons ($|X_{\pm}^{(\text{B})}|^2$), and the spectral overlap between the polariton and dark modes ($J_{\pm}$). This equation is reminiscent of the F\"orster resonance energy transfer rate \cite{Forster1948}. Note that the linear relationship between the polariton dephasing rate and the spectral overlap has been numerically demonstrated for exciton-polaritons \cite{Groenhof2019}. Similar analytic results as Eq. \eqref{eq:gamma_pm} have also been obtained previously \cite{Du2018,Saez-Blazquez2018}.
	
	Finally, given Eq. \eqref{eq:polariton_rate_simple} and Eq. \eqref{eq:gamma_pm}, the observed polariton lifetime becomes
	\begin{equation}\label{eq:polariton_rate_complete}
		\frac{1}{\tau_{\pm}} =    |X_{\pm}^{(\text{c})}|^2 k_{c} + |X_{\pm}^{(\text{B})}|^2 k_{\text{B}} + \gamma_{\pm}(k_c, k_\text{B}).
	\end{equation}
	Here, we write $\gamma_{\pm}(k_c, k_\text{B})$ (instead of $\gamma_{\pm}$) to emphasize the fact that the polaritonic dephasing rate $\gamma_{\pm}$ does depend on $k_c$ and $k_{\text{B}}$, whose values can modify $J_{\pm}$ by altering the polaritonic lineshape (see Eq. \eqref{eq:Jpm}). \footnote{Of course, $\gamma_{\pm}$ is also a function of the Rabi splitting.} In this sense, the difference between Eq. \eqref{eq:polariton_rate_complete} and Eq. \eqref{eq:polariton_rate_simple} not only comes from the introduction of the polaritonic dephasing rate $\gamma_{\pm}$ (which arises from intermolecular interactions), but also stems from the more complicated parameter dependence on $k_c$ and $k_\text{B}$, e.g., $\partial (1/\tau_{\pm})/\partial k_c = |X_{\pm}^{(c)}|^2 + \partial \gamma_{\pm} /\partial k_c \neq |X_{\pm}^{(c)}|^2$. \text{See Appendix \ref{Appendix:Nonadditive} for a detailed discussion.}

	For Eq. \eqref{eq:polariton_rate_complete}, let us consider the limit when the cavity and the bright modes are decoupled. In this limit, Eq. \eqref{eq:polariton_rate_complete} is expected to recover the lifetimes of the cavity and the bright mode, respectively. For example, in the limit of a highly red-detuned cavity mode ($\omega_c \ll \omega_0$), the "LP" is mostly composed of the cavity mode ($|X_{-}^{(\text{c})}|^2\rightarrow 1$ and $|X_{-}^{(\text{B})}|^2\rightarrow 0$) and the "UP" is mostly composed of the bright mode ($|X_{+}^{(\text{c})}|^2\rightarrow 0$ and $|X_{-}^{(\text{B})}|^2\rightarrow 1$). The lifetimes of the two "polaritons" become
	\begin{subequations}\label{eq:gamma_limit}
	    \begin{align}
	        \frac{1}{\tau_{-}} &\rightarrow k_{c}, \\
	        \frac{1}{\tau_{+}} &\rightarrow k_{\text{B}} + 2\pi \Delta^2 \int_{0}^{+\infty} d\omega \rho_0(\omega) \rho_{0}(\omega) .
	    \end{align}
	\end{subequations}
	In this limit, the "LP" lifetime is the inverse of the cavity loss rate ($k_{c}$), and the "UP" lifetime becomes the inverse of  the bright-mode relaxation ($k_{\text{B}}$) plus the dephasing rates ($2\pi \Delta^2 \int_{0}^{+\infty} d\omega \rho_0(\omega) \rho_{0}(\omega)$) outside a cavity. This expected result shows the self-consistency of Eq. \eqref{eq:polariton_rate_complete}.

	\subsection{CavMD simulations of polariton lifetime}
	Apart from the above analytic approach, the recently developed CavMD simulations  \cite{Li2020Water,Li2020Nonlinear} can also be used to determine the polariton lifetime under VSC. Within the framework of CavMD, the coupled photon-nuclear dynamics are propagated on an electronic ground-state surface. The light-matter Hamiltonian for CavMD is defined as follows:
	\begin{subequations}\label{eq:H}
		\begin{equation}
		\begin{aligned}
		\hH_\text{QED}^{\text{G}} = \ & \hH_{\text{M}}^{\text{G}} 
		+ \hH_{\text{F}}^{\text{G}} ,
		\end{aligned}
		\end{equation}
		where $\hH_{\text{M}}^{\text{G}}$ is the conventional molecular (kinetic + potential) Hamiltonian on an electronic ground-state surface outside a cavity, and $\hH_{\text{F}}^{\text{G}}$ denotes the field-related Hamiltonian:
		\begin{equation}\label{eq:H_QM}
		\begin{aligned}
		\hH_{\text{F}}^{\text{G}} & = \sum_{k,\lambda}
		\frac{\hat{\widetilde{p}}_{k, \lambda}^2}{2 m_{k, \lambda}} \\
		& + 
		\frac{1}{2} m_{k,\lambda}\omega_{k,\lambda}^2 \Bigg (
		\hat{\widetilde{q}}_{k,\lambda}  +  \frac{\varepsilon_{k,\lambda} }{m_{k,\lambda} \omega_{k,\lambda}^2}
		 \sum_{n=1}^{N} \hat{d}_{ng, \lambda}
		\Bigg )^2 .
		\end{aligned}
		\end{equation}
		Here, $\hat{\widetilde{p}}_{k, \lambda}$, $\hat{\widetilde{q}}_{k, \lambda}$, $\omega_{k,\lambda}$, and $m_{k, \lambda}$ denote the momentum operator, position operator, frequency, and the auxiliary mass for the cavity photon mode defined by a wave vector $\mathbf{k}$ and polarization direction $\vxi_\lambda$. The auxiliary mass $m_{k, \lambda}$ introduced here is solely for the convenience of molecular dynamics simulations, and the value of $m_{k, \lambda}$ does not change the VSC dynamics.\cite{Li2020Water,Li2020Nonlinear} $\hat{d}_{ng, \lambda}$ denotes the electronic ground-state dipole operator for molecule $n$ projected along the direction of $\vxi_\lambda$. The quantity $\varepsilon_{k,\lambda}\equiv \sqrt{m_{k,\lambda}\omega_{k,\lambda}^2/\Omega\epsilon_0}$	characterizes the coupling strength between each cavity photon and individual molecule, where $\Omega$ denotes the cavity mode volume and $\epsilon_0$ denotes the vacuum permittivity. 
	\end{subequations}
	
	In order to reduce the computational cost, we (i) map all quantum operators in Eq. \eqref{eq:H} to classical variables and also (ii) apply periodic boundary conditions for the molecular degrees of freedom, i.e., in Eq. \eqref{eq:H_QM} we assume $\sum_{n=1}^{N} d_{ng, \lambda} = N_{\text{cell}}\sum_{n=1}^{N_{\text{sub}}} d_{ng, \lambda}$, where $N_{\text{cell}}$ denotes the number of the periodic cells and $N_{\text{sub}} = N / N_{\text{cell}}$ is the number of molecules in a single simulation cell. By also denoting $\dbtilde{q}_{k,\lambda} = \widetilde{q}_{k,\lambda} / \sqrt{N_{\text{cell}}}$ and an effective light-matter coupling strength
	\begin{equation}
	\widetilde{\varepsilon}_{k,\lambda} \equiv \sqrt{N_{\text{cell}}} \varepsilon_{k,\lambda} = \sqrt{\frac{N_{\text{cell}} m_{k,\lambda}\omega_{k,\lambda}^2}{\Omega\epsilon_0}},
	\end{equation}
	we obtain  classical equations of motion for the coupled photon-nuclear system that are suitable for practical simulations:
	\begin{subequations}\label{eq:EOM}
	    \begin{align}
		M_{nj}\ddot{\mathbf{R}}_{nj} &= \mathbf{F}_{nj}^{(0)}  + \mathbf{F}_{nj}^{\text{cav}} + \mathbf{F}_{nj}^{\text{ext}}(t) 
		\\
		m_{k,\lambda}\ddot{\dbtilde{q}}_{k,\lambda} &= - m_{k,\lambda}\omega_{k,\lambda}^2 \dbtilde{q}_{k,\lambda}
		-\widetilde{\varepsilon}_{k,\lambda} \sum_{n=1}^{N_{\text{sub}}}d_{ng,\lambda}
	\end{align}
	\end{subequations}
	Here, the subscript $nj$ denotes the $j$-th atom in molecule $n$, and $\mathbf{F}_{nj}^{(0)}$ denotes the  force on each nucleus outside a cavity;
	\begin{equation*}
	    \mathbf{F}_{nj}^{\text{cav}} = - 
	\sum_{k,\lambda}
	\left (
	\widetilde{\varepsilon}_{k,\lambda} \dbtilde{q}_{k,\lambda}
	+ \allowbreak \frac{\widetilde{\varepsilon}_{k,\lambda}^2}{m_{k,\lambda} \omega_{k,\lambda}^2} \sum_{l=1}^{N_{\text{sub}}} d_{lg,\lambda}
	\right) 
	\frac{\partial d_{ng, \lambda}}{\partial \mathbf{R}_{nj}}
	\end{equation*}
	denotes the cavity force on each nucleus.
	$\mathbf{F}_{nj}^{\text{ext}}(t) = Q_{nj} \vE_{\text{ext}}(t)$ denotes the external force on each nucleus [with partial charge $Q_{nj}$ (due to the bare nuclear charge plus the shielding effect due to the surrounding electrons)] that arises from the time-dependent electric field ($\vE_{\text{ext}}(t)$). Note that, since our goal is to excite the polariton (which is a hybrid light-matter state), it is equivalent to either pump the cavity mode or pump the molecules given the fast energy exchange between these two subsystems (which is characterized by the Rabi splitting). In Eq. \eqref{eq:EOM}, we have chosen to excite the molecular subsystem just for simplicity [to avoid the introduction of a new phenomenological term when pumping the cavity mode is considered (namely the effective transition dipole moment of the cavity mode)]. 
	
	\subsubsection{Calculating polariton lifetime}\label{sec:method_cavmd_lifetime}
		
	The numerical scheme in Eq. \eqref{eq:EOM} allows us to calculate of the polariton lifetime by performing a nonequilibrium CavMD simulation after an IR pulse excitation. Following the method and simulation details in Ref. \cite{Li2020Nonlinear}, we capture the polariton lifetime as follows. We consider a liquid \ch{CO2} system (where $N_{\text{sub}} = 216$) in a cubic simulation cell with a cell length of 24.292 \AA (which corresponds to a molecular density of 1.101 g/cm$^3$). This molecular system is coupled to a cavity mode with two polarization directions (along $x$ and $y$); see the cartoon in Fig. \ref{fig:CavMD_summary} for the simulation setup.
	At $t =0$, the coupled cavity-molecular system has been thermally equilibrated under an NVT (constant molecular number, volume, and temperature) ensemble  with an Langevin thermostat applied to all particles (nuclei + photons). 
	If we assume no cavity loss, when $t>0$, the equilibrated system is propagated under an NVE (constant molecular number, volume, and energy) ensemble, and the polariton is pumped by sending an external $x$-polarized pulse
	\begin{equation}\label{eq:pulse}
	    \vE_{\text{ext}}(t) = E_0\cos(\omega t + \phi) \ve_x
	\end{equation}
	during a time interval $t_{\text{start}} < t < t_{\text{end}}$ ps, where the pulse frequency $\omega = \omega_{\pm}$ excites either the LP or UP and $\phi$ denotes a random phase. For parameters, we set $t_{\text{start}} = 0.1$ ps, $t_{\text{end}} = 0.6$ ps, $E_0 = 3.084 \times 10^{6}$ V/m  ($6\times 10^{-4}$ a.u.), so the input pulse fluence is $F = \frac{1}{2}\epsilon_0 c E_0^2 (t_{\text{end}} - t_{\text{start}}) = 6.32$ mJ/cm$^{2}$. 
	
	Such a pulse fluence is taken strong enough so that the polariton relaxation signal overcomes thermal and computational noise. 
	At the same time, this fluence is weak enough so that the system is not significantly excited (e.g., the magnitude of cavity excitation in Fig. \ref{fig:cavloss} is around one quantum) and the possible nonlinear absorption channel due to molecular anharmonicity is also small (see Fig. 6 in Ref. \cite{Li2020Nonlinear} for the nonlinear signal versus pulse fluence), so it is possible to compare CavMD results with the FGR rate derived from a harmonic model.  
	\red{We note that previous research \cite{Dunkelberger2019}  has suggested that one might find a contraction of the Rabi splitting at this fluence. 
	This contraction arises from depletion of the ground state population; such a contraction should be observable with a two-level (quantum) molecular model and, to a lesser extent, with an anharmonic (quantum) oscillator model.  
	In the latter case, a classical analog of depleting the ground state population obviously exists but will not be observed in our linear response calculation (Fig. \ref{fig:weak}a).\footnote{It is in principle possible to perform such a check with classical CavMD, but doing so would require a more involved calculation with two incoming pulses (pump and probe beams) and resolving such a signal would require additional effort. Quantum simulations of a Tavis--Cummings type model with anharmonic oscillators may be examined for comparison. Future work is necessary to explore this possibility.} }
	
	Immediately after the polariton pumping, we monitor the time-resolved dynamics of the cavity photon energy $E_{\text{ph}} = \sum_{\lambda=x,y} \dbtilde{p}_{\lambda}^2/2m_{c} + \frac{1}{2}m_{c}\omega_{c}^2\dbtilde{q}_{\lambda}^2$, where we have set the auxiliary photonic mass $m_c$ = 1 a.u. for simplicity. By fitting the cavity photon energy dynamics (which is averaged by 40 trajectories) with an exponential function, we obtain the  polariton lifetime. See Ref. \cite{Li2020Nonlinear} for all necessary simulation details. Note that during the above NVE simulation,  although the cavity loss has been ignored, bright-mode relaxation and dephasing processes have been included explicitly due to an explicit propagation of all molecular degrees of freedom.
	
	In order to further include the effect of cavity loss, during the nonequilibrium simulation, we attach a Langevin thermostat to the cavity mode only\cite{Li2021Collective}. Here, the friction lifetime of the Langevin thermostat defines the inverse of the cavity loss rate. The  procedure for obtaining the polariton lifetime remains the same as above.
	
	For technical details, the CavMD approach is implemented by modifying the I-PI code \cite{Kapil2019}, and the LAMMPS package \cite{Plimpton1995} is invoked to update the bare nuclear forces outside the cavity.

	\subsection{Connecting the FGR rate to CavMD simulations}
    
    We have introduced two (analytic + numerical) approaches to obtain the polariton lifetime. Both approaches have advantages and disadvantages. On the one hand, while the FGR rate  in Eqs. \eqref{eq:gamma_pm} and \eqref{eq:polariton_rate_complete} is very straightforward, this rate relies on many parameters, including the cavity loss rate, bright-mode relaxation rate, the Rabi splitting and cavity mode detuning (to determine $X_{\pm}^{\text{B}}$ and $X_{\pm}^{\text{c}}$),  the DOS of dark modes and the polaritons  and  intermolecular coupling $\Delta$ (to determine the dephasing rate). In short, the FGR rate relies heavily on the experimental input.
    
    On the other hand, despite the practical simplicity of CavMD, the polariton lifetime obtained from nonequilibrium CavMD relies heavily on the accuracy of molecular force fields. While state-of-art molecular dynamics techniques allow an accurate description of both molecular DOS and the relaxation rate \cite{Medders2015,Heidelbach1998,Kabadi2004,Kandratsenka2009}, an experimentally accurate simulation can be very nontrivial.
    
   With this background in mind, we will compare CavMD and the FGR rate as follows. With the anharmonic \ch{CO2} force field  \cite{Li2020Nonlinear} (which largely resembles the \ch{CO2} force field in Ref. \cite{Cygan2012} except for the use of an anharmonic \ch{C=O} bond potential),  we will directly obtain the polariton decay rate from nonequilibrium CavMD simulations after sending a pulse to excite the polariton; we will also obtain the  polariton and molecular DOS from equilibrium CavMD simulations. Thereafter, we will calculate the FGR rate and check for internal consistency. 
   
   Let us now introduce  how to estimate the FGR rate from equilibrium CavMD results in practice.
    
    \subsubsection{Practical way to calculate  $|X_{\pm}^{(\text{B})}|^2J_{\pm}$ from equilibrium CavMD simulations}
    
    For the FGR rate, among all parameters in Eqs. \eqref{eq:gamma_pm} and \eqref{eq:polariton_rate_complete}, the most nontrivial property related to the strong coupling is the molecule-weighted spectral overlap integral ($|X_{\pm}^{(\text{B})}|^2J_{\pm}$ in Eq. \eqref{eq:gamma_pm}). Here, we will demonstrate how to calculate this property practically. 
    
    Since the equilibrium IR absorption spectrum can be easily obtained from both experiments and equilibrium CavMD simulations,  in the expression for $J_{\pm}$, we approximate the DOS of the dark modes [$\rho_0(\omega)$] as the normalized molecular linear IR absorption lineshape outside the cavity, or $\rho_{\text{IR, outcav}}(\omega) /\mathcal{N}_0$, where $\mathcal{N}_0 = \int_{\omega_{\text{min}}}^{\omega_{\text{max}}} d\omega \rho_{\text{IR, outcav}}(\omega)$ denotes the normalization factor and 	$\omega_{\text{min}}$ and $\omega_{\text{max}}$ denote the lower and upper limits of the vibrational lineshape.  Similarly, we approximate the DOS of the LP or UP [$\rho_{\pm}(\omega)$] as the normalized linear molecular IR absorption lineshape for the polariton, respectively. Note that, because  the LP and  UP lineshapes appear together in the  molecular IR spectrum inside the cavity  [$\rho_{\text{IR, incav}}(\omega)$], here we simply assign a dividing frequency $\omega_{\text{mid}} \equiv (\omega_{-} + \omega_{+})/2$ to separate the LP and UP and obtain the lineshapes for them, respectively. Quantitatively, we can calculate $|X_{\pm}^{(\text{B})}|^2J_{\pm}$ as follows:
    \begin{subequations}\label{eq:J_pm_practical_calculation}
    	\begin{align}
    		|X_{-}^{(\text{B})}|^2J_{-} &\approx  \frac{|X_{-}^{(\text{B})}|^2}{\mathcal{N}_0 \mathcal{N}_{\text{LP}} } \int_{\omega_{\text{min}}}^{\omega_{\text{mid}}} d\omega \rho_{\text{IR, outcav}}(\omega) \rho_{\text{IR, incav}}(\omega),  \label{eq:J_pm_practical_calculation_1}\\
    		|X_{+}^{(\text{B})}|^2J_{+} &\approx  \frac{|X_{+}^{(\text{B})}|^2}{\mathcal{N}_0 \mathcal{N}_{\text{UP}}} \int_{\omega_{\text{mid}}}^{\omega_{\text{max}}} d\omega \rho_{\text{IR, outcav}}(\omega) \rho_{\text{IR, incav}}(\omega). \label{eq:J_pm_practical_calculation_2}
    	\end{align}
    	The integral in Eq. \eqref{eq:J_pm_practical_calculation_1} spans from $\omega_{\text{min}}$ to $\omega_{\text{mid}}$, which accounts for the spectral overlap between the LP and the dark modes; similarly, the integral in Eq. \eqref{eq:J_pm_practical_calculation_2} spans from $\omega_{\text{mid}}$ to $\omega_{\text{max}}$, which accounts for the spectral overlap between the UP and the dark modes. The denominators above ensure the normalization of the corresponding lineshapes:
    	\begin{align}
    		\mathcal{N}_0 &=  \int_{\omega_{\text{min}}}^{\omega_{\text{max}}} d\omega \rho_{\text{IR, outcav}}(\omega), \\
    		\mathcal{N}_{\text{LP}} &=  \int_{\omega_{\text{min}}}^{\omega_{\text{mid}}} d\omega \rho_{\text{IR, incav}}(\omega), \\
    		\mathcal{N}_{\text{UP}} &=  \int_{\omega_{\text{mid}}}^{\omega_{\text{max}}} d\omega \rho_{\text{IR, incav}}(\omega).
    	\end{align}
    \end{subequations}
    Note that, from our previous study \cite{Li2020Water}, we have learned that the integrated peak area of the LP (or UP) is proportional to the molecular weight ($|X_{\pm}^{(\text{B})}|^2$) times a correction factor (which transforms a classical dipole autocorrelation function to an experimentally comparable IR spectrum), i.e., $\int_{\omega_{\text{min}}}^{\omega_{\text{mid}}} d\omega \rho_{\text{IR, incav}}(\omega) / \int_{\omega_{\text{min}}}^{\omega_{\text{max}}} d\omega \rho_{\text{IR, incav}}(\omega) \sim  |X_{-}^{(\text{B})}|^2$ and $\int_{\omega_{\text{mid}}}^{\omega_{\text{max}}} d\omega \rho_{\text{IR, incav}}(\omega) / \int_{\omega_{\text{min}}}^{\omega_{\text{max}}} d\omega \rho_{\text{IR, incav}}(\omega) \sim  |X_{+}^{(\text{B})}|^2$ (where we have assumed that the correction factor can be largely eliminated between the numerator and denominator). Hence, we may rewrite Eq. \eqref{eq:J_pm_practical_calculation} in a simpler form:
    \begin{subequations}\label{eq:J_pm_practical_calculation_final}
    	\begin{align}
    		|X_{-}^{(\text{B})}|^2J_{-} &\approx  \frac{1}{\mathcal{N}} \int_{\omega_{\text{min}}}^{\omega_{\text{mid}}} d\omega \rho_{\text{IR, outcav}}(\omega) \rho_{\text{IR, incav}}(\omega),  \\
    		|X_{+}^{(\text{B})}|^2J_{+} &\approx  \frac{1}{\mathcal{N}} \int_{\omega_{\text{mid}}}^{\omega_{\text{max}}} d\omega \rho_{\text{IR, outcav}}(\omega) \rho_{\text{IR, incav}}(\omega),
    	\end{align}
    where
    \begin{equation}
    	\mathcal{N} = \int_{\omega_{\text{min}}}^{\omega_{\text{max}}} d\omega \rho_{\text{IR, incav}}(\omega) \int_{\omega_{\text{min}}}^{\omega_{\text{max}}} d\omega' \rho_{\text{IR, outcav}}(\omega').
    \end{equation}
    \end{subequations}	
	This expression avoids directly evaluating $|X_{\pm}^{(\text{B})}|^2$ and appears easier to handle compared with Eq. \eqref{eq:J_pm_practical_calculation}. Note that it is preferable to avoid evaluating $|X_{\pm}^{(\text{B})}|^2$ explicitly since, in CavMD, the inclusion of the self-dipole term in the Hamiltonian \eqref{eq:H} leads to a slightly different expression for $|X_{\pm}^{(\text{B})}|^2$   relative to the quantum model (where the self-dipole term is excluded --- although when the Rabi splitting is small, such difference is always negligible  \cite{Li2020Water}).
    
    By comparing the polariton lifetime fitted from nonequilibrium CavMD simulations and $|X_{\pm}^{(\text{B})}|^2J_{\pm}$ in Eq. \eqref{eq:J_pm_practical_calculation_final}, we can evaluate the inherent consistency and validity of CavMD as far as describing the polariton relaxation process.
    
    \subsection{Evaluating IR spectroscopy}
    For the sake of completeness, the equation we use to numerically evaluate the molecular IR spectrum outside the cavity is given as follows \cite{McQuarrie1976,Gaigeot2003,Habershon2013,Nitzan2006}:
    \begin{equation}\label{eq:IR_equation_outcavity}
    	\begin{aligned}
    		n(\omega)\alpha(\omega) &= \frac{\pi \beta \omega^2}{2\epsilon_0 V c} \frac{1}{2\pi}   \int_{-\infty}^{+\infty} dt \ e^{-i\omega t}  \avg{\vmu_S(0)\cdot \vmu_S(t)}.
    	\end{aligned}
    \end{equation}
    Here, $\alpha(\omega)$  denotes the absorption coefficient, $n(\omega)$ denotes  the refractive index, $V$ is the volume of the system (i.e., the simulation cell), $c$ denotes the speed of light,  $\beta = k_B T$, and $\vmu_S(t)$ denotes the \textit{total} dipole moment of the molecules at time $t$. Similarly, inside the cavity, we evaluate the IR spectrum as follows:
    \begin{equation}\label{eq:IR_equation_cavity}
    	\begin{aligned}
    		n(\omega)\alpha(\omega) &= \frac{\pi \beta \omega^2}{2\epsilon_0 V c} \frac{1}{2\pi}   \int_{-\infty}^{+\infty} dt \ e^{-i\omega t} \\
    		& \times   \avg{\sum_{i=x, y}\left(\vmu_S(0)\cdot \ve_i\right) \left(\vmu_S(t)\cdot \ve_i\right)},
    	\end{aligned}
    \end{equation}
    where $\ve_i$ denotes the unit vector along direction $i=x, y$.
    Different from Eq. \eqref{eq:IR_equation_cavity}, inside the cavity, since the supposedly $z$-oriented cavity (see the cartoon in Fig. \ref{fig:CavMD_summary}) is coupled to the molecular dipole moment along the $x$ and $y$ directions, only these two components of the molecular dipole moment are included when we calculate the IR spectrum of the polaritons.

	\section{Results and Discussion}\label{sec:results}
	
	According to the analytic expression in Eq. \eqref{eq:polariton_rate_complete}, the polariton lifetime is determined by three factors: (i) the vibrational relaxation rate of the bright mode (to the ground state) outside the cavity $k_{\text{B}}$, (ii) the cavity loss rate $k_c$, and (iii) the polariton dephasing rate to the dark modes $\gamma_{\pm}$. 
	As far as $k_{\text{B}}$ is considered, since the bright  and the dark modes are degenerate outside the cavity, the vibrational relaxation rates of the bright and dark modes should also be the same, equal to the vibrational relaxation rates of individual molecules.	For the liquid \ch{CO2} system we simulate, since the vibrational relaxation rate outside the cavity is the smallest ($\sim 0.03$ ps$^{-1}$),\cite{Li2021Collective} while the other two factors usually take a rate of $\sim$ ps$^{-1}$, below we will neglect the contribution of $k_{\text{B}}$ to the polariton lifetime and focus only on the contributions of $k_c$ and $\gamma_{\pm}$.

	\subsection{Role of cavity loss}
	
	\begin{figure*}
		\centering
		\includegraphics[width=1.0\linewidth]{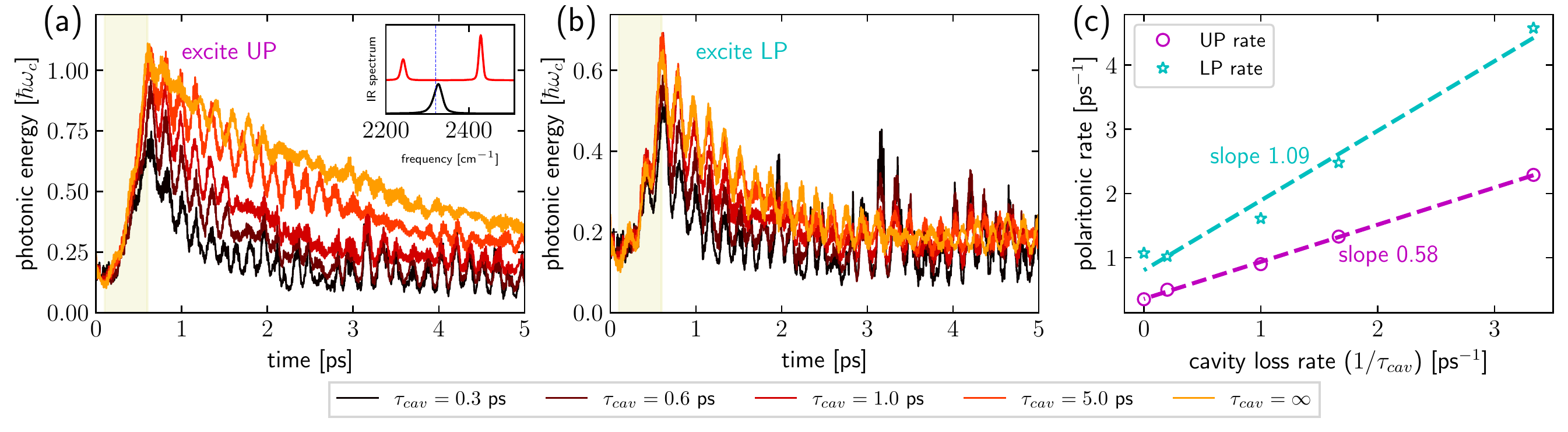}
		\caption{The polariton lifetime for liquid \ch{CO2} under VSC as a function of cavity loss ($k_c = 1/\tau_{c}$). Time-resolved cavity photon energy dynamics after a weak (a) UP or (b) LP excitation. Lines with different colors denote different cavity lifetimes (from black to yellow, the cavity lifetime spans from $\tau_c = 1/k_c = 0.3$ ps to $+\infty$; see the legend). The inset of Fig. a plots the equilibrium IR spectrum of liquid \ch{CO2} outside (black) or inside (red) the cavity, where the cavity mode frequency is 2320 cm$^{-1}$ (the vertical blue dashed line) and the effective coupling strength is  $\widetilde{\varepsilon} = 2\times 10^{-4}$ a.u. (c) The corresponding fitted polaritonic decay rate ($1/\tau_{\pm}$) versus the cavity loss rate ($k_c = 1/\tau_c$) for the UP (magenta circles) and the LP (cyan stars). The slopes of the linear fits (dashed lines) are labeled accordingly. See Sec. \ref{sec:method_cavmd_lifetime} for  other simulation details.}
		\label{fig:cavloss}
	\end{figure*}

	\subsubsection{Liquid \ch{CO2}}
	
	Let us first investigate the polariton lifetime dependence on the cavity loss rate $k_c$ when a liquid \ch{CO2} system  (with a molecular density of 1.101 g/cm$^3$) forms VSC. The Fig. \ref{fig:cavloss}a inset plots the equilibrium IR spectrum outside a cavity (black line; see Eq. \eqref{eq:IR_equation_outcavity} for the definition). Here, the \ch{C=O} asymmetric stretch of \ch{CO2} peaks at 2327 cm$^{-1}$, in good agreement with experimental values \cite{Seki2009}. When  a cavity mode (with two polarization directions) at $2320$ cm$^{-1}$ (denoted as the vertical blue line in the inset of Fig. \ref{fig:cavloss}a) is coupled to the molecular subsystem with an effective coupling strength $\widetilde{\varepsilon} = 2\times 10^{-4}$ a.u., a pair of UP (peaked at $\omega_{+}=2428$ cm$^{-1}$) and LP (peaked at $\omega_{-}= 2241$ cm$^{-1}$)  is observed in the molecular IR spectrum (red line; see Eq. \eqref{eq:IR_equation_cavity} for the definition).
	
	Under  VSC conditions, Fig. \ref{fig:cavloss}a plots the photonic energy dynamics when the UP is resonantly pumped by a relatively weak external E-field [$\vE_{\text{ext}}(t) = E_0\cos(\omega t + \phi) \ve_x$ with fluence 6.32 mJ/cm$^2$ and a 0.5 ps duration, which is labeled as a yellow shadow]; see Sec. \ref{sec:method_cavmd_lifetime} for  details on the IR excitation. When the cavity lifetime ($\tau_{c} = 1/k_c$) is tuned from $\infty$ (orange line) to 0.3 ps (black line), the photonic energy decays faster. In Fig. \ref{fig:cavloss}a, the fast oscillations (with a period of $\sim 0.2$ ps) during the photonic energy decay is in agreement with the Rabi splitting (187 cm$^{-1}$ = 0.18 ps$^{-1}$),  indicating a fast coherent energy exchange between cavity photons and the \ch{C=O} asymmetric stretch.  Note that for Fabry--P\'erot VSC, the experimental cavity lifetime usually takes $1 \sim 10$ ps \cite{Xiang2020Science}.
	
	Since cavity photons contribute  to the polaritons, we can extract the polariton lifetime ($1/\tau_{\pm}$) by fitting the photonic energy dynamics with an exponential function after the IR excitation (when $t > 0.6$ ps). Fig. \ref{fig:cavloss}c plots the fitted UP lifetime ($1/\tau_{+}$) versus the cavity loss rate (magenta circles). A linear fit of the UP lifetime data yields a slope of $0.58$, meaning that roughly half of the cavity loss rate contributes to the polariton decay.
	
	After investigating the UP lifetime, we study the LP lifetime dependence on the cavity loss rate. Similarly, Fig. \ref{fig:cavloss}b plots the photonic energy dynamics when the LP is resonantly excited by the external E-field with the same fluence (6.32 mJ/cm$^{2}$) as the UP excitation. Fig. \ref{fig:cavloss}c  plots the fitted LP decay rate versus the cavity loss rate (cyan stars), where the  slope of the LP data is 1.09 after a linear fit.

	At this moment, we cannot make an immediate connection between the CavMD results with the analytic rate in Eq. \eqref{eq:polariton_rate_complete}. As emphasized below Eq. \eqref{eq:polariton_rate_complete} (see also Appendix \ref{Appendix:Nonadditive}), since changing the value of $k_c$ not only directly modifies $|X_{\pm}^{(\text{c})}|^2$, but also indirectly alters the polaritonic dephasing rate $\gamma_{\pm}$ through the polaritonic lineshape, the slope obtained  by fitting $1/\tau_{\pm}$ versus $k_c$ (as in Fig. \ref{fig:cavloss}c) does not necessarily  recover $|X_{\pm}^{(\text{c})}|^2$--- it is $|X_{\pm}^{(\text{c})}|^2 + \partial \gamma_{\pm} / \partial k_c$ instead. Therefore, only when $\gamma_{\pm}$  is not very sensitive to $k_c$ ($|\partial \gamma_{\pm} / \partial k_c| \ll |X_{\pm}^{(\text{c})}|^2$), we expect to obtain the agreement between the fitted slope and $|X_{\pm}^{(\text{c})}|^2$. In our case, the cavity mode and the \ch{C=O} asymmetric stretch are near resonance, so  the cavity weight of the polariton  ($|X_{\pm}^{(\text{c})}|^2$) is roughly a half. In Fig. \ref{fig:cavloss}c, we have observed that the fitted slope for the UP (not the LP) agrees well with $|X_{+}^{(\text{c})}|^2$. Because the value of $\partial \gamma_{\pm} / \partial k_c$  is unknown, however, the agreement between the UP fitted slope and $|X_{+}^{(\text{c})}|^2$ cannot fully verify any consistency between CavMD and the analytic rate, not to mention the LP data.
	
	\subsubsection{A Diluted \ch{CO2} ensemble}
	
	\begin{figure*}
		\centering
		\includegraphics[width=1.0\linewidth]{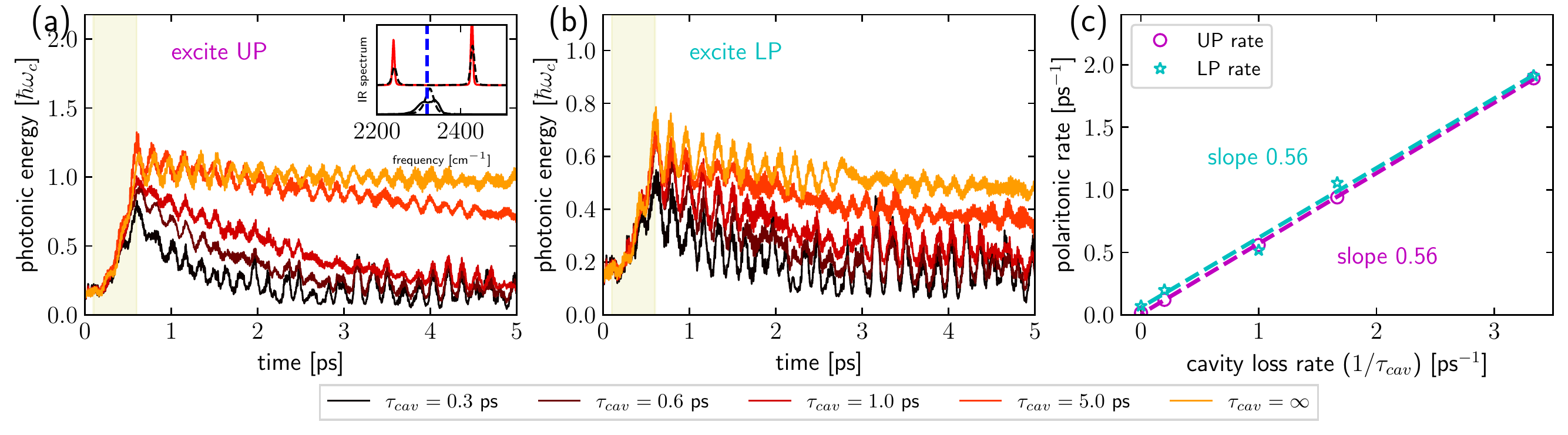}
		\caption{The same plot as Fig. \ref{fig:cavloss} when the \ch{CO2} molecular density is decreased from 1.101 g/cm$^3$ to 0.073 g/cm$^3$ and all  other simulation details are kept the same. In Fig. (a) inset the dashed  black lines are the corresponding spectra in Fig. \ref{fig:cavloss} and the solid lines are the spectra when the density is reduced. As shown in the inset,  the Rabi splitting is largely unchanged when the molecular density  changes.}
		\label{fig:cavloss_dilute}
	\end{figure*}

	In order to better compare the CavMD results with the analytic expression in Eq. \eqref{eq:polariton_rate_complete}, we need to exclude the influence of polaritonic dephasing in the CavMD results (note that $k_{\text{B}}$ is always negligibly small for our simulations). This can be achieved, 
	since $\gamma_{\pm} \propto \Delta^2$ (Eq. \eqref{eq:gamma_pm}), by reducing intermolecular interactions ($\Delta$) or equivalently decreasing  the \ch{CO2} molecular density.  Hence, we further repeat our simulation in Fig. \ref{fig:cavloss} with a diluted \ch{CO2} ensemble (with a density of 0.073 g/cm$^3$). The molecular density is reduced by increasing the simulation cell length to 60.0 \AA  while fixing $N_{\text{sub}}$, the molecular number in the cell, so the Rabi splitting remains the same (see the inset of Fig. \ref{fig:cavloss_dilute}a). One can imagine such a process can be realized experimentally by changing the thickness of the \ch{SO2} layers between the cavity mirrors (see Fig. \ref{fig:CavMD_summary} middle for the cavity setup) so that the volume containing the molecules can be adjusted with the fixed cavity length (i.e., with fixed light-matter coupling per molecule and Rabi splitting). 
	
	keeping all the other simulation details the same as Fig. \ref{fig:cavloss}, in Fig. \ref{fig:cavloss_dilute} we replot the photonic energy dynamics for the (a) UP and (b) LP as well as (c) the fitted polaritonic decay rates versus the cavity loss rate. As shown in Fig. \ref{fig:cavloss_dilute}c, the fitted slopes for both the UP and LP are nearly a half, in agreement with $|X_{\pm}^{\text{c}}|^2$.  Because here we have greatly suppressed the polaritonic dephasing rate by decreasing the molecular density, Fig. \ref{fig:cavloss_dilute}c confirms the consistency between CavMD and the analytic rate in Eq. \eqref{eq:polariton_rate_complete} when the cavity loss dependence is considered.

	As a short summary of this part, we have observed that, if the polaritonic dephasing rate is  negligible, both  CavMD and the analytic expression (Eq. \eqref{eq:polariton_rate_complete}) predict a similar dependence for the polaritonic decay rate as a function of the cavity loss, i.e., $|X_{\pm}^{(\text{c})}|^2$ of the cavity loss rate enters the polaritonic decay rate. While the fitted slope of the UP is always $|X_{\pm}^{(\text{c})}|^2$ times the cavity loss rate for both Figs. \ref{fig:cavloss} and \ref{fig:cavloss_dilute}, the LP results for liquid \ch{CO2} (see Fig. \ref{fig:cavloss}) no longer shows this dependence on the cavity loss. The deviation for the LP is an example for the complicated interplay between cavity loss  and polaritonic dephasing under liquid-phase VSC, which might prevent the LP from forming a more coherent and long-lived state than the cavity photon mode.
	

	\subsection{Role of polariton dephasing to the dark modes}
	
	Next, let us focus on the effect of polariton dephasing on the polariton lifetime. For the  results below, we set the cavity loss rate to be zero. Since the bright-mode relaxation rate $k_B$ is much smaller than the dephasing rate, below the fitted polariton decay rate from CavMD simulations should largely equal to the polaritonic dephasing rate.

	\begin{figure*}
		\centering
		\includegraphics[width=1.0\linewidth]{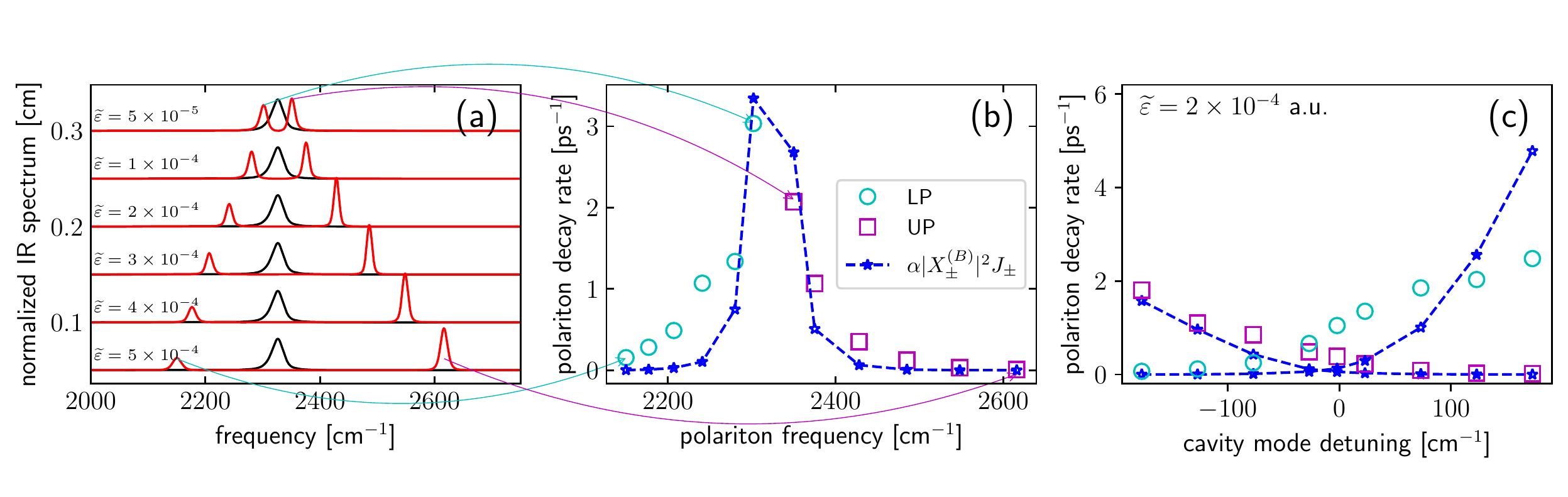}
		\caption{Polariton decay rate versus Rabi splitting and cavity mode detuning. (a) Simulated IR spectrum for liquid \ch{CO2} inside (red lines) or outside (black lines) a cavity. The cavity mode frequency ($\omega_c$) is set as 2320 cm$^{-1}$. From top to bottom, the effective coupling strength is labeled on each lineshape (from $5\times 10^{-5}$ a.u. to $5\times 10^{-4}$ a.u.), respectively. (b) Simulated polariton decay rate for  the polaritons in Fig. a (cyan circles for the LP and magenta squares for the UP; same as below). Blue stars plot $\alpha|X_{\pm}^{(\text{B})}|^2 J_{\pm}$, where $|X_{\pm}^{(\text{B})}|^2 J_{\pm}$ is defined in Eq. \eqref{eq:J_pm_practical_calculation_final} and $\alpha = 1200 \text{\ ps}^{-1}/\text{cm}$ is a prefactor to best fit the polaritonic decay rates.  (c) Polariton decay rate and the corresponding $\alpha|X_{\pm}^{(\text{B})}|^2 J_{\pm}$ versus the cavity mode detuning ($\omega_c - \omega_0$) when $\widetilde{\varepsilon} = 2\times 10^{-4}$ a.u., where $\alpha$ takes the same value as that in Fig. b. See Sec. \ref{sec:method_cavmd_lifetime} for  other simulation details.}
		\label{fig:weak}
	\end{figure*}

	\subsubsection{Rabi splitting dependence}
	Fig. \ref{fig:weak}a plots the equilibrium IR spectrum calculated from CavMD simulations. Here, the cavity mode frequency is $\omega_c = 2320$ cm$^{-1}$, and the effective light-matter coupling strength is set from $\widetilde{\varepsilon} = 5\times 10^{-5}$ a.u. to $5\times 10^{-4}$ a.u. (red lines from  bottom to top; see the text on each lineshape which labels the corresponding $\widetilde{\varepsilon}$). We have also plotted the equilibrium IR spectrum outside the cavity (black lines) for comparison.
	
	For every polariton shown in Fig. \ref{fig:weak}a, Fig. \ref{fig:weak}b  plots the polariton dephasing rate from nonequilibrium CavMD simulations when a weak pulse (with fluence 6.32 mJ/cm$^2$) resonantly excites the corresponding polariton.  Here, the fitted LP (cyan circles) and UP (magenta squares) decay rates are plotted as a function of the polaritonic peak frequency (see Fig. \ref{fig:weak}a). As mentioned above, because the bright-mode relaxation rate $k_{\text{B}}$ is intrinsically small and we have also set the cavity loss rate to zero, the fitted polariton lifetime from nonequilibrium CavMD simulations  should capture the polariton dephasing rate.

	In order to explicitly compare the simulated polaritonic dephasing rate with the FGR calculation, Fig. \ref{fig:weak}b also plots the molecule-weighted spectral overlap ($|X_{\pm}^{(\text{B})}|^2 J_{\pm}$; blue stars) for the polaritons by numerically evaluating Eq. \eqref{eq:J_pm_practical_calculation_final} given the lineshapes in Fig. \ref{fig:weak}a. Here, in order to directly compare with the polariton dephasing rate from CavMD simulations, we multiple a prefactor $\alpha = 1200 \text{\ ps}^{-1}/\text{cm}$ on top of $|X_{\pm}^{(\text{B})}|^2 J_{\pm}$, where the prefactor $\alpha$ is chosen to best fit the CavMD dephasing rates. In Fig. \ref{fig:weak}b, the  $\alpha|X_{\pm}^{(\text{B})}|^2 J_{\pm}$ values and the fitted rates generally agree with each other (especially for the UP). Both  the UP and LP results show a maximum when the polariton frequency is closer to the \ch{C=O} asymmetric stretch outside the cavity ($\omega_0 = 2327$ cm$^{-1}$). However, the agreement disappears for the LP regime near 2241 cm$^{-1}$, where the CavMD rates are larger than the  $\alpha|X_{\pm}^{(\text{B})}|^2 J_{\pm}$ values. Such a disagreement can be explained by the polariton-enhanced molecular nonlinear absorption mechanism \cite{Li2020Nonlinear}: when the LP frequency is close to  2241 cm$^{-1}$ and twice the LP is near resonance with $0\rightarrow 2$ vibrational transition of the dark modes, an additional nonlinear dephasing mechanism occurs for the LP and causes a faster LP dephasing rate than that from a harmonic model ($|X_{\pm}^{(\text{B})}|^2 J_{\pm}$). Because the pulse fluence we use here is not very strong, the nonlinear effect does not dominate the polariton dephasing, and only a modest disagreement between the fitted rates and $|X_{\pm}^{(\text{B})}|^2 J_{\pm}$ appears for  the LP regime.  Note that in Fig. 6 of Ref. \cite{Li2020Nonlinear} we have shown that, when the LP frequency is  2241 cm$^{-1}$, after the LP excitation, the  high excited state population of the dark modes amplifies nonlinearly when the pulse fluence increases, affirming our suggestion of a nonlinear effect.

	As a side note, let us mention that  the rotating wave approximation used in our quantum model is expected to fail in the ultrastrong coupling limit \cite{FriskKockum2019}, i.e., when $\Omega_N/2\omega_0 > 0.1$  (which corresponds to a Rabi splitting of $\Omega_N\sim 400$ cm$^{-1}$ for our case).  In Fig. \ref{fig:weak}, only a single data point falls in this regime ($\widetilde{\varepsilon} = 5\times 10^{-4}$ a.u.) and we believe the results here should be valid.
	
	\subsubsection{Detuning dependence}
	
	After studying how vibrational relaxation and polaritonic dephasing depends on the Rabi splitting, we will now address how these observables depend on  the cavity mode frequency detuning.  When we set the effective coupling strength to $\widetilde{\varepsilon} = 2\times 10^{-4}$ a.u. and tune the cavity mode frequency  from  2150 cm$^{-1}$ to 2500 cm$^{-1}$, Fig. \ref{fig:weak}c plots the fitted polaritonic dephasing rate (magenta squares for the UP and cyan circles for the LP) versus the cavity mode detuning from nonequilibrium CavMD simulations after a weak, resonant pumping of the corresponding polaritons. Similar to Fig. \ref{fig:weak}b, here we also compare the fitted rate to  $\alpha|X_{\pm}^{(\text{B})}|^2 J_{\pm}$ (where $\alpha = 1200 \text{\ ps}^{-1}/\text{cm}$ is set the same as Fig. \ref{fig:weak}b), which is calculated from the linear IR spectra inside versus outside the cavity (see Eq. \eqref{eq:J_pm_practical_calculation_final}). Note that the IR spectra for a detuned cavity mode has been plotted in Ref. \cite{Li2020Nonlinear} and is not given here. 
	
	As shown in Fig. \ref{fig:weak}c, the two results agree well for the UP (magenta squares). For the LP, there is a modest underestimation of  $\alpha|X_{\pm}^{(\text{B})}|^2 J_{\pm}$  compared with the  fitted dephasing rate (cyan circles) when the cavity mode detuning is near zero, which corresponds to the LP frequency ($\sim 2241$ cm$^{-1}$) where  polariton enhanced molecular nonlinear absorption  occurs. Interestingly, when the cavity mode is very positively detuned,  the fitted dephasing rate becomes slightly smaller than  $\alpha|X_{\pm}^{(\text{B})}|^2 J_{\pm}$, where we expect that the two results agree with each other  (since the nonlinear absorption mechanism should vanish when the LP frequency becomes closer to the fundamental vibrational frequency). For the moment, there is no obvious explanation for this disagreement. Because the fitted dephasing rate is very large when the cavity mode is very positively detuned, an error in numerical fitting can not be entirely ruled out.
		
	\subsubsection{Molecular density dependence}
	
	\begin{figure*}
		\centering
		\includegraphics[width=1.0\linewidth]{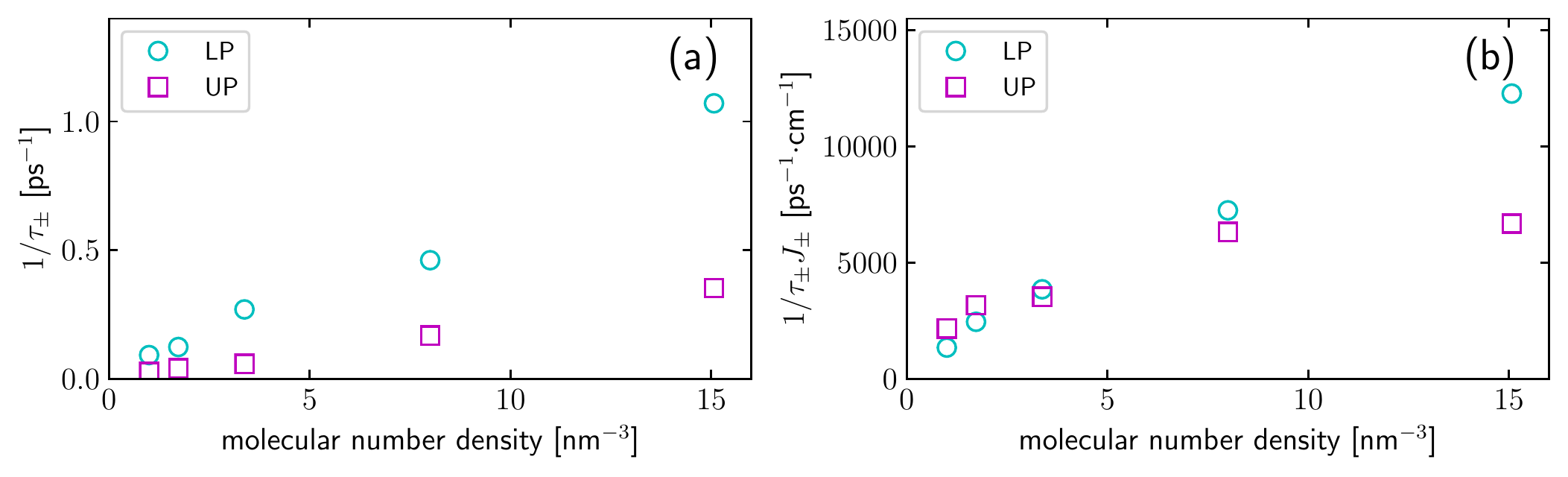}
		\caption{(a) Polariton decay rate ($1/\tau_{\pm}$) versus the molecular number density when  $\omega_c = 2320$ cm$^{-1}$ and $\widetilde{\varepsilon} = 2\times 10^{-4}$ a.u. Here, the molecular density is changed by tuning the simulation cell length while keeping the molecular number the same (so the Rabi splitting is fixed). See Sec. \ref{sec:method_cavmd_lifetime} for  other simulation details. (b) The corresponding $1/(\tau_{\pm}J_{\pm}$) versus the molecular number density. }
		\label{fig:weak_density}
	\end{figure*}
	
	The Rabi splitting and cavity mode detuning studies have shown that the fitted polaritonic dephasing rate agrees with  $|X_{\pm}^{(\text{B})}|^2 J_{\pm}$ very well if polariton-enhanced molecular nonlinear mechanism does not play an important role. In order to fully validate the agreement between CavMD and the FGR rate in Eq. \eqref{eq:polariton_rate_complete}, we further
	examine the correlation between intermolecular interactions ($\Delta$) and the polaritonic dephasing rate. Figs. \ref{fig:cavloss} and \ref{fig:cavloss_dilute} have shown such an example: by decreasing intermolecular interactions (via reducing the molecular density), both the UP and LP lifetimes become longer. 
	
	Fig. \ref{fig:weak_density} shows more comprehensive data examining how the polariton decay rate ($1/\tau_{\pm}$) depends on molecular density (under a fixed Rabi splitting). When the cavity mode frequency is  $\omega_{c} = 2320$ cm$^{-1}$ and the effective light-matter coupling is $\widetilde{\varepsilon} = 2\times 10^{-4}$ a.u., Fig. \ref{fig:weak_density}a shows that both the fitted UP and LP dephasing rates increase when the molecular density increases. Moreover,  this dependence appears to be linear. Note that as shown in the inset of Fig. \ref{fig:cavloss_dilute}a, since changing the molecular density can slightly modify the lineshapes (while keeping the Rabi splitting constant), the spectral overlap $J_{\pm}$ is also a function of the molecular density. 
	In order to exclude the influence of $J_{\pm}$, in Fig. \ref{fig:weak_density}b we plot the corresponding $1/(\tau_{\pm}J_{\pm})$ versus the molecular density. Again,  $1/(\tau_{\pm}J_{\pm})$ increases when the molecular density increases. This dependence highlights the fact that it is intermolecular interactions which control the polariton dephasing rate under a weak external excitation; this conclusion is consistent with the premise of  the FGR rate in Eq. \eqref{eq:polariton_rate_complete}, although the linear dependence on the density no longer holds for the UP.
	
	\subsubsection{Molecular system size dependence}
	
	There is a piece of hidden information in Eq. \eqref{eq:polariton_rate_complete}: the polaritonic dephasing rate should be independent of the molecular number --- provided the  Rabi splitting, cavity mode frequency, and molecular density are the same.  In Ref. \cite{Li2020Nonlinear} (see Fig. 7 therein), we have reported that the CavMD polariton dephasing rate shows this independence for both the LP and the UP.

	\section{Conclusion}\label{sec:conclusion}
	
	In conclusion, we have extensively studied the parameter dependence (including the cavity loss, Rabi splitting, cavity mode detuning, and molecular density) of the polariton relaxation lifetime with nonequilibrium CavMD simulations when the polariton is excited with a weak laser pulse. By comparing the simulation results with a FGR rate from a tight-binding harmonic model, we find a very good agreement for the UP lifetime and  sometimes a modest disagreement for the LP. Such a disagreement does not void the validity of CavMD. By contrast, it is an indicator of the incompleteness of the quantum model we use: this model does not consider the effect of molecular anharmonicity on the nature of highly excited vibrational states, and so this model fails to capture polariton enhanced molecular nonlinear absorption \cite{Li2020Nonlinear,Ribeiro2020}, an important mechanism which can also alter the  relaxation of the polaritons. Please see Ref. \cite{Li2020Nonlinear} for a detailed analysis on the LP-induced molecular nonlinear absorption phenomena using CavMD.
	
	 In fact, the agreement between nonequilibrium CavMD simulations and the FGR rate is not surprising: on the one hand, classical molecular dynamics simulations can often describe molecular vibrational relaxation outside the cavity even for the high frequency vibrations \cite{Heidelbach1998,Kabadi2004,Kandratsenka2009}; on the other hand, in the area of light-matter interactions, classical theory has been shown to agree with quantum FGR rates very well (at least as far as parameter dependence), e.g.,  the spontaneous emission rate inside \cite{Krasnok2015} or outside \cite{Jackson1999,Miller1978} a cavity. Hence, by inheriting the both sides, CavMD is expected to deal with many VSC dynamics reasonably well.
	
	As far as the polariton relaxation mechanism is considered, our simulation indicates that since the polariton inherits the photonic character only partially and the formation of Rabi splitting separates the lineshapes between molecular bright and dark modes, the polariton lifetime can be longer than both the cavity lifetime and the bright-mode dephasing lifetime outside the cavity (where the bright and dark modes are degenerate and the dephasing rate should be very large). Thus, polaritons (especially the UP) may exhibit a more long-lived coherence than either the cavity photon or the molecular subsystem \cite{Grafton2021}, demonstrating the benefits of forming strong light-matter interactions. By designing cavities with a high Q factor (or small loss) and  a very large Rabi splitting, it is possible to create molecular polaritons which could survive for a very long time before relaxation. Such long-lived molecular polaritons might be a good platform for quantum information applications in the future.
	
	Finally, we emphasize the advantages and limitations of using classical CavMD simulations on describing VSC-related phenomena. As  summarized in Fig. \ref{fig:CavMD_summary} and the surrounding text, CavMD is an affordable tool which can recover many nonreactive VSC dynamics when realistic molecules are considered. Hence, from a practical point of view, for an arbitrary VSC-related problem, due to the low cost of CavMD, it is always beneficial to run a simulation and check if an experimental finding can be roughly obtained from this approach. However, we must keep in mind that as a classical simulation, CavMD  is expected to provide only a qualitative description of VSC dynamics and some experimental results cannot be quantitatively described by CavMD. For example, for pump-probe and 2D-IR experiments of VSC, the exact frequency-resolved nonlinear spectra \cite{Dunkelberger2016,Dunkelberger2018,Dunkelberger2019,Xiang2018,Xiang2019Nonlinear,Xiang2019State,Xiang2020Science,Grafton2021,Xiang2021} cannot be recovered from a direct nonequilibrium classical CavMD simulation and one must do some quantum or semiclassical treatments to recover the frequency-resolved information in the nonlinear regime. Looking forward, there are many research opportunities within the framework of CavMD, including improving the performance on frequency-resolved nonlinear spectra and extending this approach to simulate quantum effects and chemical reactions.

	
	\begin{appendices}
		
		\setcounter{equation}{0}
		\setcounter{table}{0}
		\renewcommand{\theequation}{S\arabic{equation}}
		\renewcommand{\bibnumfmt}[1]{[S#1]}
		\setcounter{section}{0}
		\renewcommand{\thesection}{S-\Roman{section}}
		
		\section{The pulse profile}\label{Appendix:Efield}	
		\begin{figure}[h]
			\centering
			\includegraphics[width=1.0\linewidth]{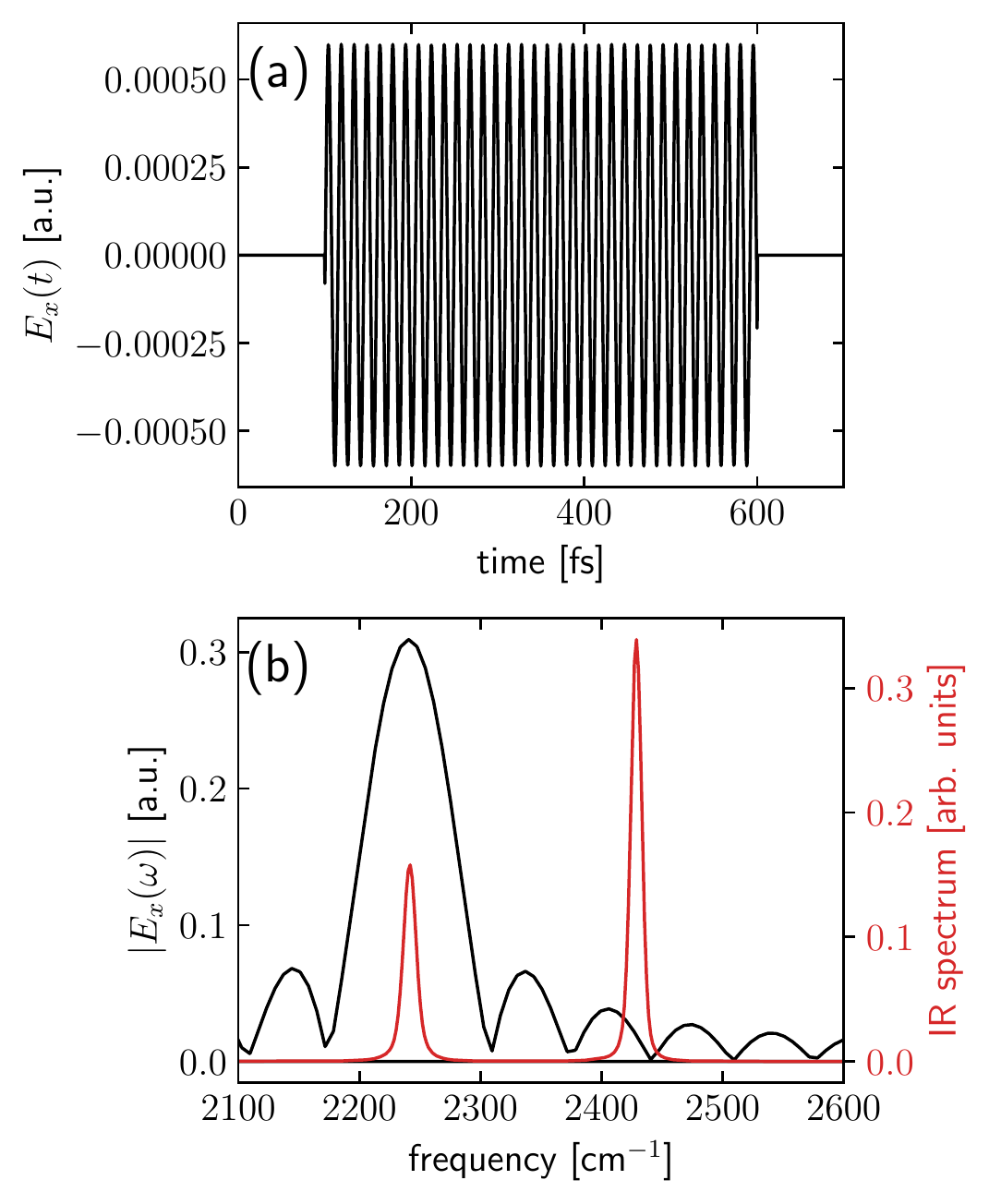}
			\caption{(a) The E-field profile $E_x(t)$ defined in Eq. \eqref{eq:pulse}, where the pulse frequency is set as the LP frequency of Fig. \ref{fig:cavloss}: $\omega = \omega_{\text{LP}} =2241$ cm$^{-1}$. (b) The corresponding E-field spectrum in the frequency domain. The red line replots the Rabi splitting spectrum in the inset of Fig. \ref{fig:cavloss}a.}
			\label{fig:Efield}
		\end{figure}
	
	Fig. \ref{fig:Efield} plots the profile of the pulse [Eq. \ref{eq:pulse}] in both time and frequency domains.

	\section{Nonadditive effects in polaritonic decay rates}\label{Appendix:Nonadditive}	
	
	Polaritons are hybrid light-matter states, so one might naively think that a polaritonic property is a linear combination of the photonic contribution (with a weight $|X_{\pm}^{(c)}|^2$) plus the molecular bright-mode contribution (with a weight $|X_{\pm}^{(\text{B}})|^2$). For example, for the polaritonic decay rate under VSC, if one applies Eq. \eqref{eq:polariton_rate_simple}, one can simply obtain:
	\begin{subequations}
		\begin{align}
			\frac{\partial (1/\tau_{\pm}) }{\partial k_c}
			&=  |X_{\pm}^{(c)}|^2 \\
			\frac{\partial (1/\tau_{\pm})}{\partial k_\text{\text{B}}} &= |X_{\pm}^{(\text{B})}|^2 
		\end{align}
	\end{subequations}
	
	In this work, we have found the importance of the polaritonic dephasing rate ($\gamma_{\pm}$) on the polaritonic decay rate. According to Eqs. \eqref{eq:gamma_pm} and \eqref{eq:polariton_rate_complete}, since $\gamma_{\pm}$ is a complicated function of $k_c$, $k_{\text{B}}$, and the Rabi splitting, we find that
	\begin{subequations}
		\begin{align}
			\frac{\partial (1/\tau_{\pm}) }{\partial k_c}
			&= |X_{\pm}^{(c)}|^2 + \frac{\partial \gamma_{\pm}}{\partial k_c} \neq |X_{\pm}^{(c)}|^2 \\
			\frac{\partial (1/\tau_{\pm})}{\partial k_\text{\text{B}}} &= |X_{\pm}^{(\text{B})}|^2 + \frac{\partial \gamma_{\pm}}{\partial k_{\text{B}}} \neq |X_{\pm}^{(\text{B})}|^2 
		\end{align}
	\end{subequations}
	This equation implies that the simple additive feature no longer holds when the polaritonic dephasing rate is large. For example, increasing the cavity loss rate $k_c$ can broaden the polaritonic lineshape and alter the spectral overlap between the polaritons and dark modes, leading to a modified $\gamma_{\pm}$. Similarly, changing $k_{\text{B}}$ can simultaneously alter the polaritonic and dark-mode lineshapes, which can also influence $\gamma_{\pm}$. In Fig. \ref{fig:cavloss}c, we have shown that $\partial (1/\tau_{-}) /\partial k_c \neq |X_{\pm}^{(c)}|^2$ for liquid \ch{CO2} under VSC.
		
	\end{appendices}

	\section{Acknowledgments}
	This material is based upon work supported by the U.S. National Science Foundation under Grant No. CHE1953701 (A.N.); 
	and US Department of Energy, Office of Science,
	Basic Energy Sciences, Chemical Sciences, Geosciences, and Biosciences Division under Award No. DE-SC0019397  (J.E.S.).
	
	\section{DATA AVAILABILITY STATEMENT}
	The data that support the findings of this study
	(including source code, input and post-processing
	scripts, and the corresponding tutorials) are openly
	available on Github at \url{https://github.com/TaoELi/
	cavity-md-ipi} \cite{TELi2020Github}.

	%

\end{document}